\documentclass[letterpaper,prb,twocolumn,groupedaddress]{revtex4}
\usepackage{graphicx}
\usepackage{calc}
\usepackage{pifont}
\usepackage{amssymb}
\usepackage{epsfig}
\usepackage{amssymb}
\usepackage{amsmath}
\usepackage{color}

\DeclareMathAlphabet{\mathitb}{OT1}{cmr}{bx}{sl}
\begin{document}

\renewcommand{\thefootnote}{\fnsymbol{footnote}}
\title{Kondo Resonances in Molecular Devices}
\author{G. D. Scott$^1$}
\email{gavin.scott@rice.edu}
\author{D. Natelson$^{1,2}$}
\affiliation{$^1$Department of Physics and Astronomy, $^2$ Department of Electrical and Computer Engineering, Rice University, 6100 Main St., Houston, TX 77005\\}

\date{\today}

\begin{abstract}

Molecular electronic devices currently serve as a platform for studying a variety of physical phenomena only accessible at the nanometer scale.  One such phenomenon is the highly correlated electronic state responsible for the Kondo effect, manifested here as a ``Kondo resonance'' in the conductance.  Because the Kondo effect results from strong electron-electron interactions, it is not captured by the usual quantum chemistry approaches traditionally applied to understand chemical electron transfer.  In this review we will discuss the origins and phenomenology of Kondo resonances observed in single molecule devices, focusing primarily on the spin-$\frac{1}{2}$ Kondo state arising from a single unpaired electron.  We explore the rich physical system of a single-molecule device, which offers a unique spectroscopic tool for investigating the interplay of emergent Kondo behavior and such properties as molecular orbital transitions and vibrational modes.  We will additionally address more exotic systems, such as higher spin states in the Kondo regime, and we will review recent experimental advances in the ability to manipulate and exert control over these nanoscale devices.
\\

\end{abstract}

\maketitle

Research in the field of molecular electronics has been driven by an assortment of promising technological advancements, both fundamental and applied in nature.  The discrete building blocks used to construct modern electronic components will ultimately limit the extent to which these devices can be scaled down in size.  Device miniaturization may be taken to new extremes by fabricating components
that utilize a single molecule or molecular complex as the active element.  The realization of such devices has presented a challenge to our ability to manipulate and exert control over these nanoscale structures, and has provided new insights and an expanded understanding of the way in which a molecule may interact with solid state components.

Early theories of electrical conduction through molecules were presented in the 1940's by Robert Mulliken and Albert Szent-Gyorgi.  However, a practical foundation for the premise of a device operating with a single, custom designed molecule is widely credited as having begun with the theories of Aviram and Ratner, put forth in 1974.\cite{Aviram1974}  They proposed using a donor-acceptor molecule as a molecular diode.  Shortly thereafter came an upsurge of papers presenting more detailed analyses of molecular conduction and designs for molecular wires, switches, and logic circuits.  The elementary
structure of all of these devices, regardless of the application, consists of a molecule or molecular complex attached to two metallic electrodes functioning as source and drain contacts.  The incorporation of a third, capacitively coupled electrode, typically referred to as a gate, allows for the formation of a 3-terminal transistor-like device.  This structure is then analogous to the field effect transistor (FET), the basis for nearly all modern electronic devices.

The experimental realization of even the most basic of such designs first required the development of new fabrication techniques enabling means of contacting individual molecules \emph{via} metallic electrodes.  These techniques led to early transport studies, primarily directed at investigating conduction through molecules \emph{via} two terminal $I-V$ measurements.  The field has taken sizable steps forward recently thanks to collaborations between chemists
synthesizing tailor-made complexes, and physicists incorporating these complexes into nanoscale devices.

This review focuses primarily on Kondo physics in single-molecule electronic devices.  As we explain in more detail below, the Kondo effect, first noticed in the 1930s in bulk metals containing magnetic impurities, is a consequence of strong electron-electron interactions and correlations.  It therefore lies outside the scope of single-particle quantum chemistry descriptions, while profoundly affecting electron transport.  The Kondo effect leads to a pronounced zero bias Kondo resonance in the conduction that would be absent in a single-particle picture.  We first introduce the fabrication techniques required to produce single-molecule electronic devices.  After an overview of the physics of electron transport in such nanodevices, we discuss the origin of the Kondo effect.  Implementations of the Kondo effect in molecules are presented together with the means by which these systems can be probed in order to better understand the correlation effects.   We also discuss the way in which experimental observations deviate from commonly used models of the Kondo state.  Finally, we address scenarios beyond the single channel spin-1/2 system and possible future lines of investigation.\\
\\
\textbf{FABRICATION}\\

Single molecule devices, serving here as a platform for the study of the Kondo effect, are often described as a double barrier tunneling structure in which a molecule is connected to metallic source and drain electrodes by contacts that each act as tunneling barriers.  This arrangement can be accomplished by more than one method.

\begin{figure}
\centering
\includegraphics [scale=.25]{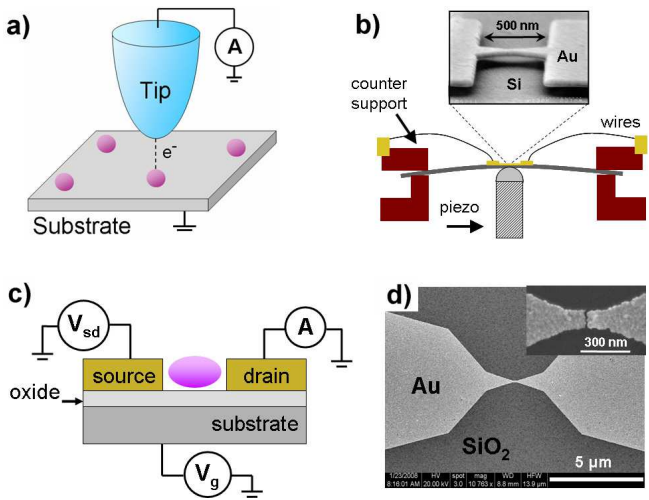}\\
\caption{\textbf{a)} Schematic of STM measurement setup.  \textbf{b)} Schematic of mechanically controlled break junction apparatus.  Top: SEM image of Au nanowire suspended 40 nm above the substrate. (Adapted from Ref. 6 with permission from the American Physical Society). \textbf{c)} Idealized single molecule break junction device configuration with capacitively coupled gate electrode.  \textbf{d)} SEM image of lithographically defined Au nanowire defined by e-beam lithography on a Si/SiO$_2$ substrate.  Inset: higher resolution image of similar device after electromigration has created a nanometer-size gap.}
\label{Fabrication1}
\end{figure}

In a scanning tunneling microscope (STM), the tunneling current is measured between a sharp metallic tip (capable of controlled movement in the $x$, $y$, and $z$ directions) and a conducting surface.  This tool enables one to map the density of states (DOS) of a surface covered with dispersed molecules.  When a molecule of interest is found, situating the molecule between the sharp tip and the conducting substrate establishes a double barrier junction (Fig. \textbf{\ref{Fabrication1}a}).  Electrons can tunnel between the substrate (including the molecule) and the atom on the STM tip that is situated closest to the substrate surface. The absence of a gate electrode limits the STM to two terminal measurements.  Kondo physics in STM experiments has been observed for more than a decade,\cite{Madhavan1998,Li1998} with much recent progress being made in molecular systems, as described below.

Alternately, planar single-molecule devices may be fabricated using
``breakjunction'' techniques.  Arrays of nanowires are lithographically defined on an insulated substrate.  A dilute solution of the molecules of interest is spread on the substrate surface, with molecules adhering either through physisorption or the formation of a self-assembled monolayer.  The nanowires are then broken in a controlled manner, either through a mechanical process or \emph{via} electromigration.  Ideally, after the breaking process a molecule will be situated in the newly formed gap. The broken ends of the wire can then act as source and drain electrodes while a capacitively coupled conducting substrate can function as a gate electrode (Fig. \textbf{\ref{Fabrication1}c}).

The tunneling probability decreases exponentially with distance between the source and drain, ensuring that interelectrode conduction is dominated by an extremely small volume, comparable to that of a single molecule, situated at the point of closest interelectrode separation.  One can assume that only a fraction of the devices that are broken will have an appropriately sized gap containing an individual active element.  It follows that the formation of a working sample using any breakjunction technique is highly statistical in nature, and many nanowires are often required to achieve an operational device.

In the mechanical breakjunction technique a post is positioned against the back of the substrate beneath the nanowires, while the substrate edges are rigidly held in place. The post is slowly extended by means of a piezoelectric stack.  The substrate bends as the piezo stack is extended, which results in the nanowire being stretched until a small break forms  (Fig. \textbf{\ref{Fabrication1}b}).\cite{Zhou1995,Champagne2005}  The large geometric reduction ratio between the movement of the piezo and the nanowire assures a high degree of control over the resulting gap size.  An alternate method is referred to as the electromigration technique,\cite{HPark1999} and has become the most widely utilized means of fabricating three-terminal single molecule transistor (SMT) devices.\cite{HPark2000,JPark2002,Liang2002,Yu2004b}  A current is ramped between the source and drain electrodes of a nanowire.  When the power dissipated reaches a certain point a nanometer size gap opens up at the narrowest portion of the nanowire through the process of electromigration and Joule heating  (Fig. \textbf{\ref{Fabrication1}d}).\\
\\
\noindent
\textbf{BASIC TRANSPORT IN SINGLE MOLECULE DEVICES}\\

Establishing that transport is through a single molecule of interest has presented a challenge to experimentalists investigating these devices. A STM produces a map of the topography and the DOS of a surface whereby one can visually search the substrate surface for an isolated molecule.  A high quality STM may reveal a detailed picture of molecular structure sufficient for unequivocally verifying the relevant molecular species.\cite{Lu2004}

The situation in break junctions is more difficult.  Whether or not conduction in a breakjunction device is occurring through a single molecule of the desired species is usually ascertained through the analysis of transport properties, as discussed at length in several references.\cite{Yu2004d,Natelson2006,Ward2008}  A number of characteristics indicative of, or antithetic to, conduction through a single molecule are evaluated before a more in-depth analysis is attempted.  In addition to determining that charge is passing through only one molecule, one may search for attributes in the transport data that uniquely identify the particular molecule of interest.\cite{Ward2008}

The primary tool for characterizing an individual device is the measurement of the differential conductance, $dI/dV$, as a function of source-drain bias, $V_{sd}$, calculated from $I-V$ traces or measured directly using standard low frequency lock-in techniques.  The dependence of this conduction data on temperature, $T$, and, in three-terminal devices, on gate voltage, $V_{g}$, provides a wealth of information about conduction mechanisms.

It is important to consider the relationship between conductance in these solid-state situations and the usual discussion of electron transfer theory in chemistry.\cite{Marcus1964,Hush1961,Newton1991}  Here the source and drain electrodes are analogous to donors and acceptors.  Furthermore, in the usual solution-based chemistry case, fluctuations in the local environment (\textit{e.g.} rearrangements of polar molecules or nearby ions) are critical to the reaction energy landscape, whereas in the solid state limit there are
no such fluctuations.  Instead, electrostatic interactions between the molecule and a solution environment {\AA}ngstroms away are replaced here by interactions with the electrodes.

\begin{figure}
\centering
\includegraphics [scale=.46]{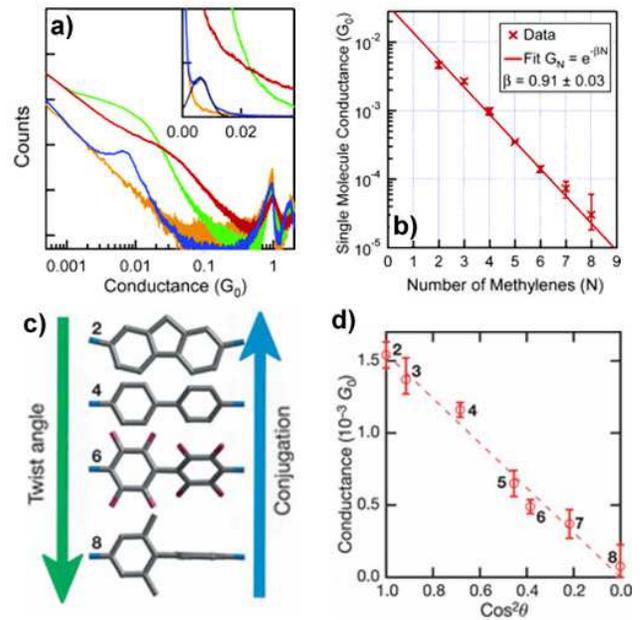}\\
\caption{\textbf{a)} Conductance histograms measured as junctions are pulled apart with 25 mV bias in the presence of 1,4-benzenediamine (blue), 1,4-benzenedithiol (red), 1,4-benzenediisonitrile (green), and without molecules (yellow) shown on a log-log plot.  Inset: same data on a linear plot with a Gaussian fit to the peak (black).  Amine terminated molecules result in the most well defined conductance.  \textbf{b)} Off-resonant conductance is measured through alkanediamines with different numbers of methylene groups in the chain.  The $\times$'s denote the center peak position in the conductance histograms plotted against the number of methylene groups in the alkane chain.  Junction conductance decreases with an increasing number of methylene groups.  (Reproduced from Ref. 19 with permission from the American Chemical Society).  \textbf{c)}  Subset of the biphenyl series studied, shown in order of increasing twist angle or decreasing conjugation.  Conductance histograms are obtained from measurements through the series of molecules with internal twist angles increasing from molecule 2 through 8.  \textbf{d)} Data points indicate histogram peak position for each molecule plotted against cos$^2\theta$, where $\theta$ is the twist angle calculated for each molecule.  The planar conformation has the highest conductance.  (Reproduced from Ref. 20 with permission from the Nature Publishing Group).}
\label{BondingPic}
\end{figure}

The tunneling barriers between the electrodes and the active element are defined by the geometry and chemistry of the molecule-electrode interfaces.  The end groups used to form chemical bonds to metallic electrodes have been a subject of particular interest as they have a direct impact on the molecule/source and molecule/drain coupling strengths, respectively labeled $\Gamma_S$ and $\Gamma_D$.  In an experiment using a modified STM configuration, current was recorded at a fixed bias voltage while repeatedly forming and breaking Au point contacts in the presence of molecules.  Comparing conductance histograms comprised of many conductance traces measured in the presence of thiol, isonitrile, and amine terminated aromatics, Venkataraman \textit{et al}. found that the choice of end-group greatly affected the distribution and the average values of measured off-resonant conductance through the junctions (Fig. \textbf{\ref{BondingPic}a}).\cite{Venkataraman2006a}  More reproducible conductance values point to a narrower distribution of end-group/electrode bonding geometries.  For example, amines are found to bind preferentially to undercoordinated Au structures.

The most elementary models treat the two contacts between the chemical end groups and the metal as isolated tunnel barriers, with low-energy accessible states in the molecule through which sequential tunneling can occur.  While this can be a useful representation at times, as discussed below, such an interpretation would not take into account the detailed molecular structure of a particular active element, which can play an important role in the resulting transport characteristics.  For instance, in studies of off-resonant conduction, a molecule's length\cite{Liang2002,Venkataraman2006a} (Fig. \textbf{\ref{BondingPic}b}) as well as its conformation\cite{Venkataraman2006b} (e.g. the relative twist angle between two rings of a $\pi$-conjugated biphenyl molecule)(Fig. \textbf{\ref{BondingPic}c,d}) exhibit distinct relationships with measurements of the average junction conductance.  It is also generally accepted that in such off-resonant transport the delocalization of the molecular orbitals make conjugated molecules more conducting than typically insulating saturated molecules.

\begin{figure}
\centering
\includegraphics [scale=.265]{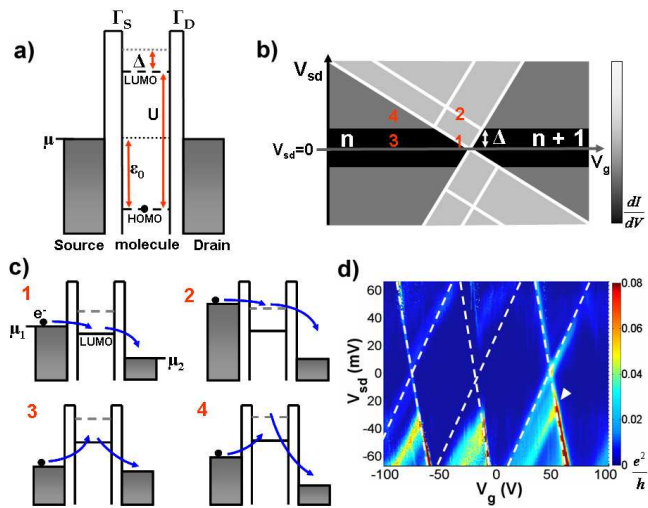}\\
\caption{\textbf{a)} Energy level diagram of double barrier tunneling junction.  \textbf{b)} Portion of an idealized stability diagram of a single molecule device displaying $dI/dV_{sd}$ (shown as relative brightness) as a function of $V_{sd}$ and $V_g$.  \textbf{c)} Features visible in the stability diagram corresponding to: (1) Resonant tunneling, (2) Resonant tunneling through on excited state, (3) Elastic cotunneling, (4) Inelastic cotunneling.  \textbf{d)} Data from a real C$_{60}$-based single molecule transistor device exhibiting some of these features.  Dashed lines indicate blockaded regions (\textit{i.e.} Coulomb diamonds).  White arrow indicates excitation corresponding to the 33 meV intra-cage vibrational mode of C$_{60}$.}
\label{BasicTransport}
\end{figure}

In the case of resonant conduction, it is often appropriate to model the system as a double barrier tunneling junction, in which the molecule is treated as a quantum dot and the details of its structure are largely ignored.  In this limit it follows that the electrostatic charging energy of a single-molecule device is non-negligible due to the small size and correspondingly small capacitance of the active element.  An additional electron is blocked from tunneling onto the molecule until this charging energy, or on-site repulsion ($U$), is overcome.  This suppressed conduction is known as the Coulomb blockade.  The ``particle-in-a-box'' spacing between discrete single-particle energy levels in a molecule, $\Delta$, is also non-negligible.  At equilibrium the spacing between single particle energy states is the energy spacing between the highest occupied molecular orbital level (HOMO level) and the lowest unoccupied molecule orbital level (LUMO level), commonly referred to as the HOMO-LUMO gap (Fig. \textbf{\ref{BasicTransport}a}).

When no molecular state is available between the respective Fermi levels of the source and drain ($\mu_1$ and $\mu_2$), there is no resonant tunneling, and current is said to be blockaded. At sufficiently large $V_{sd} = (\mu_{1}-\mu_{2})/e$, an accessible energy level is shifted into the energy range between $\mu_{1}$ and $\mu_{2}$, overcoming the charging energy and the HOMO-LUMO gap, and allowing sequential resonant tunneling through the molecule.
This leads to a peak in $dI/dV$.

In the limit of comparatively weak molecule-metal electronic coupling ($\Gamma_{S},~\Gamma_{D} \ll \Delta$), a capacitively coupled gate electrode can shift the discrete molecular levels energetically relative to $\mu_{1}$ and $\mu_{2}$.  Thus gate bias, $V_{g}$, shifts the range of $V_{sd}$ over which current is blockaded.  Traces of $dI/dV$ measured as a function of $V_{sd}$ over a range of incrementally changing values of $V_g$ may be compiled to form a ``stability diagram'', illustrating where in the $V_{sd}-V_g$ parameter space, and with what relative magnitude, conduction is allowed (Fig. \textbf{\ref{BasicTransport}b,d}).  The resulting diamond pattern (Coulomb diamonds) shows regions of suppressed tunneling, and therefore regions of stable charge states, with fixed average molecular charge.  When $V_g$ brings a discrete level into alignment with $\mu_{1}=\mu_{2}$, resonant tunneling may occur at zero bias.  This point in the stability diagram where the blockade vanishes is referred to as a ``charge degeneracy point''.  In chemistry language, charge degeneracy points are the $V_{g}$ values where redox transitions between molecular charge states take place.

Further analysis of the differential conductance measurements can be used to uniquely identify the (species of) active element.  These features include the number of charge states (\emph{i.e.} Coulomb diamonds) accessible by application of a gate voltage; the spacing between charge state transitions (\textit{i.e.} distance between adjacent Coulomb diamonds), calculated from the maximum magnitude of gate bias required to add or remove one electron from the molecule; the size of the electron addition energy, determined from the maximum magnitude of source-drain bias required to overcome the blockaded region; the number of characteristic tunneling threshold slopes bordering blockaded regions; and the number and spacing of energy levels corresponding to excited states.

Sequential tunneling processes are also possible in which the electron tunnels elastically \emph{via} an excited state of the molecule, and in which the electrons loses energy due to tunneling events involving excited states of the molecule.The signature of this is the presence of additional $dI/dV$ peaks that run parallel to the edges of the blockaded region in stability diagrams.  Excited states of a molecule may originate from both local vibrational modes and excited electronic states of the system.  The energy of these excitations can be determined from the source-drain voltage at which they intercept the conductance gap (Fig. \textbf{\ref{BasicTransport}b-d}).  If these quantized excitations can be correlated with known molecular properties, they may serve as a fingerprint for molecular identification.

Higher order tunneling events are another common means of transport in single molecule devices.  Cotunneling, for instance, may occur in single molecule devices at bias voltages that do not overcome the Coulomb repulsion.  This becomes more apparent when the coupling between the dot and leads is increased.  During elastic cotunneling an electron tunnels into and out of the same energy level, leaving the molecule in its ground state (Fig. \textbf{\ref{BasicTransport}b,c}).  This is the process that Marcus/Hush theory refers to as superexchange.  In this off-resonant limit the molecule acts as an effective tunneling barrier.\cite{Ratner1990}

Inelastic cotunneling refers to the process by which an electron tunnels onto and off of the molecule through two different energy levels, leaving the molecule in an excited state.  Inelastic cotunneling may occur when the source-drain bias is such that $eV_{sd} = \Delta$, where $\Delta$ is the energy level spacing between the ground state and the first excited state (Fig. \textbf{\ref{BasicTransport}b,c}).\cite{Franceschi2001,Sukhorukov2001}  Inelastic cotunneling is the process responsible for inelastic electron tunneling spectroscopy (IETS).\cite{Jaklevic1966,Lambe1968}

In the next section, we discuss another higher order tunneling process involving spin that produces a zero bias resonance observed in some single-molecule devices.  This resonance is associated with the formation of a highly correlated electron state known as the Kondo effect.\\
\\
\noindent
\textbf{THE SPIN-$\frac{1}{2}$ KONDO EFFECT}\\

\begin{figure}
\centering
\includegraphics [scale=.29]{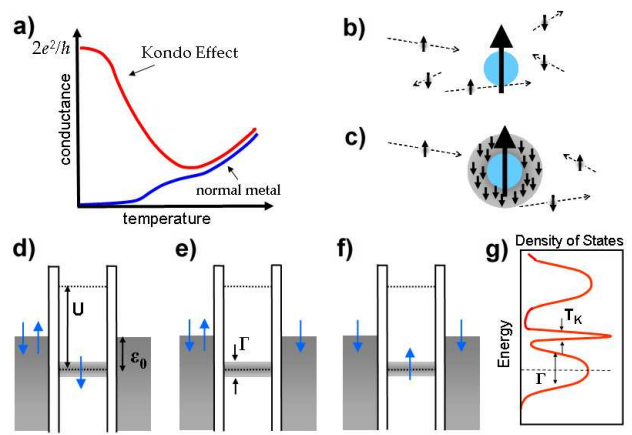}\\
\caption{\textbf{a)} Conductance \emph{vs.} temperature through a double barrier tunneling junction with and without the Kondo effect.  \textbf{b)} Magnetic impurity in a sea of conduction electrons above the Kondo temperature $T_K$.  \textbf{c)} Formation of the Kondo screening cloud when $T < T_K$.  Kondo transport mechanism:  \textbf{d)} initial state, \textbf{e)} virtual state (during which time an electron can hop off of the impurity), \textbf{f)} final state.  The replacement electron of the final state can be of either spin orientation.  \textbf{g)} Density of states as a function of the equilibrium Fermi level.  (Adapted from Ref. 38 with permission from The Institute of Physics.)}
\label{KondoTransport}
\end{figure}

At room temperature the electrical resistance of a typical metal is dominated by electron-phonon scattering.  As $T$ is reduced, this resistance decreases since electrons can travel more easily as the lattice vibrations of the metallic crystal decrease.  Neglecting superconductivity, as $T \rightarrow 0$ the resistance for most metals is expected to saturate to a constant value, with electron scattering set by residual impurities and structural defects.  It was discovered in the 1930s that some metals instead exhibit an increase in resistance at low temperatures (Fig. \textbf{\ref{KondoTransport}a}).\cite{Haas1934}  It was subsequently shown that this low temperature resistance increase is correlated with the presence of magnetic impurities.\cite{Sarachik1964}  In 1964 Jun Kondo presented a theory that was able to successfully account for this resistance anomaly, which today we know as the Kondo effect.\cite{Kondo1964}  The increased low temperature resistance was shown to originate from an increase in the effective scattering cross-section of the magnetic impurities.  A cloud of conduction electrons around the impurities serves to screen the impurity magnetic moments and increase the scattering of electrons near the Fermi level (Fig. \textbf{\ref{KondoTransport}b,c}).

The ground state of the (impurity + conduction electrons) system is a many-body spin singlet.  At its root the Kondo effect is a consequence of the Pauli exclusion principle and strong electron-electron repulsion.  For a spin-1/2 impurity, absent interactions, a conduction electron could sit on the impurity, pairing up with the spin of the local moment, thereby forming a singlet.  Including interactions, the strong Coulomb repulsion makes this energetically prohibitive.  However, at low temperature higher order tunneling processes may allow the delocalized conduction electrons to develop a polarization that compensates for the localized magnetic moment.  The quenching of the magnetic moment by these higher order processes is what serves to ``screen'' the unpaired impurity spin and enhance the scattering of the conduction electrons while forming a new many-body state with total spin 0.

The Kondo effect is inherently a many-body phenomenon that can appear when there is some localized, non-zero spin degree of freedom coupled to a sea of conduction electrons.  Single-particle approaches alone cannot account for the existence of this effect.  Using perturbation methods, Kondo showed that the antiferromagnetic interaction between the Fermi sea and the magnetic impurities leads to the observed logarithmic rise in electron-impurity scattering as temperature is reduced.  While self-consistent density functional theory methods have shown accurate predictions for some properties of molecular junctions, these \textit{ab-initio} approaches fail to capture the strong electron correlation effects leading to the conductance resonance associated with the Kondo state.\cite{Wang2008,Neel2007}  A complete physical description requires non-perturbative methods, namely Renormalization Group (RG).\cite{Gruner1974,Wilson1975,Nozieres1980,Andrei1983}  Numerical RG techniques were later developed that could accurately calculate transport properties, such as resistivity, over a broad range of temperatures.\cite{Hewson1993,Costi1994}

An idealization of this system is the Anderson single-level impurity model, where the magnetic impurity is viewed as having one energy level with an unpaired electron at energy level $\varepsilon_0$ (Fig. \textbf{\ref{KondoTransport}d}).\cite{Anderson1961} All energy levels below $\varepsilon_0$ are fully occupied with electrons, and all energy levels above
$\varepsilon_0$ are unoccupied.  Neither the levels above nor below $\varepsilon_0$ contribute to the Kondo resonance.  When $\varepsilon_0$ is below the equilibrium Fermi level, the unpaired electron is trapped with spin-1/2.  It is classically forbidden to bring the electron out of the impurity without adding energy into the system.  However, the uncertainty principle does allow for this to happen \emph{via} an exchange process whereby the electron is allowed to jump off the magnetic impurity and onto an electrode on a time scale $\sim
h/|\varepsilon_0|$ (Fig. \textbf{\ref{KondoTransport}e}).  During this time, another electron from the Fermi sea must jump onto the impurity to replace the one that has left (Fig. \textbf{\ref{KondoTransport}f}).  The spin of this new electron may be either spin-up or spin-down.  The coherent superposition of many such co-tunneling events results in the screening of the local spin, thereby producing the Kondo resonance (Fig. \textbf{\ref{KondoTransport}g}).  In a bulk metal such a resonance scatters electrons with energies near the Fermi level.  Because the same electrons are responsible for the low-temperature conductivity, this scattering leads to an increase in resistance. In the single-molecule device configurations relevant to us, all electrons must travel through the active element, and transport is dominated by tunneling through a single magnetic site.  The enhanced scattering in this case is forward scattering that couples states from the source and drain electrodes, leading to a corresponding increase in conductance as $T \rightarrow 0$ (Fig. \textbf{\ref{KondoTransport}a}).\cite{Appelbaum1966,Ng1988,Costi1994}

The state giving rise to the Kondo effect, consisting of a localized degenerate state coupled to an electron reservoir, is well characterized by the Anderson Hamiltonian,\cite{Anderson1961}
\begin{eqnarray}
H_{Anderson} = \sum_{k\sigma}{\varepsilon}_kc^\dag_{k\sigma}c_{k\sigma} + \sum_\sigma{\varepsilon}_{\sigma}d^\dag_{\sigma}d_\sigma + Un_{d_{\uparrow}}n_{d_\downarrow}\nonumber\\
+ \sum_{k\sigma}{(\nu_{k}d^\dag_\sigma{c_{k\sigma}} + \nu_k^\ast{c^\dag_{k\sigma}}d_\sigma)}.
\label{Hamiltonian1}
\end{eqnarray}
The first two terms account for the electrons in the leads and dot, respectively, where $c^\dag_{k\sigma}$ ($c_{k\sigma}$) creates (destroys) an electron in the leads with momentum $k$, spin $\sigma$, and energy $\varepsilon_{k}$.  The third term represents the Coulomb interaction of the electrons on the island, and the last term describes the tunneling between the dot and the leads with amplitude $\nu_k$.

With a change of variables, the Schrieffer-Wolff unitary transformation can be performed on Eq.~(\ref{Hamiltonian1}).  Retaining terms up to second order in $\nu_k$, while eliminating $\nu_k$ to first order, one can arrive at the Kondo Hamiltonian, in which only the singly-occupied state is allowed on the dot.
\begin{equation}
H^{}_{Kondo} = H^{}_{cond} + H^{}_{int}.
\label{Hamiltonian2}
\end{equation}
The second term of Eq. (\ref{Hamiltonian2}) describes the interaction between the conduction electrons and the local moment as an antiferromagnetic coupling,\cite{Kondo1964,Schrieffer1966,Pustilnik2004}
\begin{equation}
H^{}_{\emph{int}} = J{\bf s^{}}_{cond}\cdot{{\bf S^{}}_{molecule}}.
\label{Hamiltonian3}
\end{equation}
where $\mathbf{S}$ is the net spin of the quantum dot, $\mathbf{s}$ is sum of spin operators of the conduction electrons, and $J$ represents the strength of the effective spin-spin interaction.

The binding energy of the singlet state (with total spin = 0) formed between the delocalized conduction electrons and the localized magnetic moment is not simply proportional to $J$.  The characteristic energy scale associated with the (continuous) transition into and out of the Kondo state is given by the Haldane relation,\cite{Haldane1978}
\begin{equation}
k^{}_BT^{}_K = \frac{\sqrt{{\Gamma}U}}{2}e^{-\pi\varepsilon_0(-\varepsilon_0 + U)/{\Gamma}U},
\label{Haldane}
\end{equation}
which gives the Kondo temperature for the Anderson model, where $U$ is the on-site repulsion and $\varepsilon_0$ is the energy level of the molecule through which transport occurs, as depicted in Fig. \textbf{\ref{KondoTransport}d}.  Thus $T_K$ is exponentially dependent on $\Gamma$, the sum of the widths of the tunneling barriers, which is in turn extremely sensitive to the precise molecule-electrode coupling.  We then find that the Kondo energy scale (and the associated zero bias resonance) will be most evident in devices that possess relatively strong molecule-electrode couplings, which also leads to increased overall conductance.

\begin{figure}
\centering
\includegraphics [scale=.28]{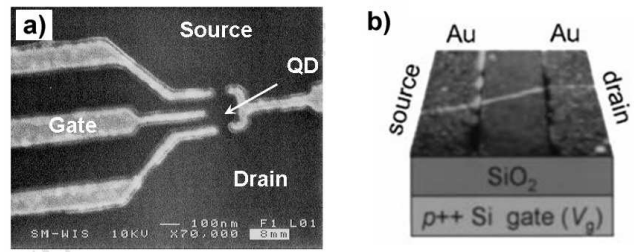}\\
\caption{Two devices used to observe the Kondo effect.  \textbf{a)} Scanning electron micrograph of an electrostatically defined GaAs/AlGaAs quantum dot. \textbf{b)} Atomic force microscope image of a device consisting of a 2 nm thick nanotube bundle with 30 nm thick gold contacts separated by 300 nm.  Sketched beneath the image are the 300 nm thick SiO$_2$ layer and a highly p-doped Si substrate.  ((a) and (b) reproduced from Refs. 44 and 54, respectively, with permission from the Nature Publishing Company).}
\label{SemiconductorQD2}
\end{figure}

Renewed interest in Kondo physics took place in the late 1990s after the first experimental demonstration of the single impurity Kondo effect was accomplished
using an electrostatically defined semiconductor quantum dot (QD), in which the conduction electrons interact with a localized ``puddle'' of electrons comprising the magnetic impurity (Fig. \textbf{\ref{SemiconductorQD2}a}).\cite{Goldhaber1998a,Cronenwett1998}  A zero bias Kondo resonance was observed due to the screening of the local moment by the conduction electrons.  At around the same time, in 1998 evidence of the Kondo effect was seen in conductance measurements through single magnetic adatoms on a conducting surface using a scanning tunneling microscope (STM).\cite{Madhavan1998,Li1998}

\begin{figure}
\centering
\includegraphics [scale=.335]{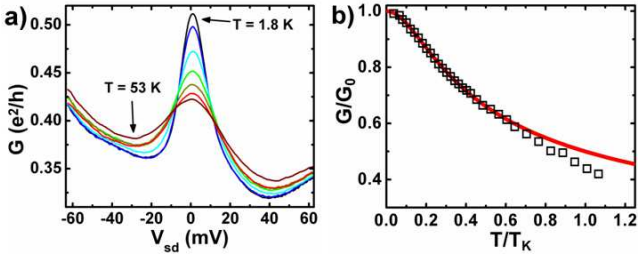}\\
\caption{\textbf{a)} Traces of $dI/dV$ \emph{vs.} $V_{sd}$ for several temperatures ranging from 1.8 K (tallest/black) to 53 K (shortest/maroon) for a single molecule device constructed \emph{via} the electromigration technique.  Amplitude of Kondo resonance decreases with increasing temperature.  \textbf{b)} Fit of equilibrium conductance \emph{vs.} temperature, normalized with respect to $G_0$ and $T_K$, respectively, using Eq. \ref{GEK}, yielding $T_K$ $\approx$ 52 K.  (Adapted from Ref. 50).}
\label{GEKpic}
\end{figure}

The general behavior of the zero bias conductance  amplitude in a quantum dot can be shown to obey a logarithmic temperature dependence.  Goldhaber-Gordon \textit{et al.} developed an empirical form derived from a fit to the renormalization group analysis,\cite{Wilson1975,Goldhaber1998b} which more accurately captures the temperature dependence of the equilibrium conductance, and can be written as
\begin{equation}
G(T,0) = \frac{G^{}_0}{(1 + (2^{1/s} - 1)(T/T_K)^2)^s} + G_b,
\label{GEK}
\end{equation}
where $s = 0.22$ for a spin-1/2 impurity, $G_0$ is the peak conductance minus $G_b$ in the limit that $T\rightarrow 0 K$, $G_0 = G(0,0) - G_b$, and $T_K$ is defined as the value of $T$ for which $G(T,0) - G_b = G_0/2$ (Fig. \textbf{\ref{GEKpic}}).  Note that in STM and single-molecule junction experiments, there can be an additional ``background'' contribution to the conductance, $G_b$, due to non-resonant processes such as direct tunneling between the source and drain electrodes.  The relative source-drain coupling can be deduced from the magnitude of the resonance peak as the temperature approaches 0~K using the equation \cite{Beenakker1991}
\begin{equation}
G_0 - G_b = \frac{2e^2}{h}\frac{(4\Gamma_{\mathrm S}\Gamma_{\mathrm D})}{(\Gamma_{\mathrm S}+\Gamma_{\mathrm D})^2}\mathit{f}(T/T_K),
\label{Asymmetry}
\end{equation}
where $\mathit{f}(T/T_K) = (1 + (2^{1/s} - 1)(T/T_K)^2)^{-s}$.

Strong correlations between conduction electrons and a single Ti atom on a Ag(100) surface leading to the Kondo effect were observed using a STM apparatus and used to study the temperature dependent broadening of the associated zero bias resonance.\cite{Nagaoka2002}  Approximating the Kondo resonance peak as a Lorentzian, the full width at half max (FWHM) has a temperature dependence that can be approximated as
\begin{equation}
FWHM = \frac{2}{e}\sqrt{{({\pi}k_BT})^2 + 2(k_BT_K)^2},
\label{FWHM}
\end{equation}
where $k_B$ is the Boltzmann constant.  The system parameters, $T_K$ and $G_0$, can then be extracted using Eqs. (\ref{GEK}) and (\ref{FWHM}).  As the temperature is increased, higher order transport processes become relevant, and deviations from these forms are expected.  The enhanced $V_{sd}$ = 0 V conductance of the Kondo resonance rapidly decreases as a function of applied bias.  Quantum interference between Kondo processes and non-Kondo processes (\emph{e.g.} direct tunneling) can modify the observed resonance in $dI/dV$ \emph{vs.} $V_{sd}$, resulting in a Fano lineshape.  Depending on the relative complex amplitudes for the processes involved, these lineshapes can take the form of peaks, dips, or intermediate asymmetric structures.\cite{Nagaoka2002,Calvo2009}

\begin{figure}[t]
\centering
\includegraphics [scale=.38]{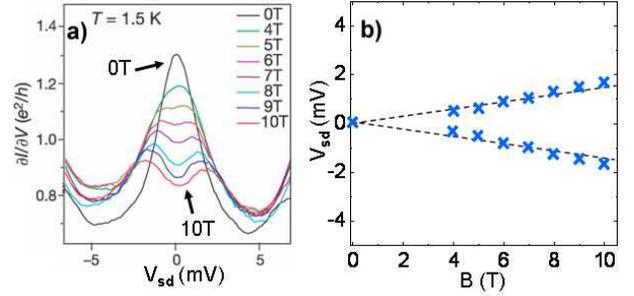}\\
\caption{\textbf{a)} Magnetic-field dependence of the Kondo resonance.  (Reproduced from Ref. 9 with permission from the Nature Publishing Company). \textbf{b)} Peak positions as a function of applied field (blue $\times$'s).  The peak splitting varies linearly with magnetic field greater than a critical value.\cite{Costi2000,Houck2005}  Dashed lines represents the predicted Zeeman splitting.}
\label{FieldSplitting}
\end{figure}

The equilibrium conductance of the resonance peak can be suppressed by an applied magnetic field which lifts the degeneracy of the up and down spin states through which transport occurs.  In the presence of an applied $B$-field, the enhanced conductance can be recovered through the application of a finite bias thereby revealing a splitting of the original Kondo resonance peak (Fig. \textbf{\ref{FieldSplitting}}).  The split peaks are expected to separate by an amount equal to twice the Zeeman energy\cite{Meir1993,Liang2002} for applied fields larger than a critical value, $B_c \sim 0.5T_K$.\cite{Costi2000}  These descriptions of the temperature and $B$-field dependence of the Kondo resonance are commonly used to verify a resonance peak originating from a single spin-1/2 impurity.

The energy scales relevant to the Kondo effect in a semiconductor QD (\emph{i.e.} $T_K$ and $\Delta$) can be prohibitively low, making observation of this phenomenon difficult to achieve.  Common values of $T_K$ in these structures, for instance are in the few hundred milliKelvin range.  This is caused by the relatively weak confinement of the electrons in the dot, which produces a larger effective dot (\emph{i.e.} impurity) size compared to what is found in a single molecule device.  The larger size of the dot leads to a correspondingly smaller charging energy, U, which by the Haldane relation, Eq.~(\ref{Haldane}), results in a smaller $T_K$.  The single impurity Kondo effect can also be observed in carbon nanotubes devices (Fig. \textbf{\ref{SemiconductorQD2}b}),\cite{Nygard2000,Paaske2006} and SMTs \cite{JPark2002,Liang2002,Yu2004b} when an unpaired electron exists on the nanotube or molecule, respectively, at temperatures below the mean energy level spacing of the device.  In carbon nanotubes the relatively greater confinement, compared to that of a semiconductor QD, has led to observed Kondo temperatures typically around a few Kelvin.  Single molecule-based devices most often exhibit Kondo temperatures in the tens to hundreds of Kelvin range.  $T_K$ in single molecule devices exceeding 100~K was first reported by Yu \textit{et al.} in 2004 using C$_{60}$ molecules.\cite{Yu2004b}  This large magnitude of $T_K$ is more readily accessible, making molecules a desirable system for studying Kondo physics.

Kondo resonances in molecular devices were first observed in some of the earliest studies of transport in single molecule break junction structures.
Park \textit{et al.} demonstrated conduction through two different molecules ([Co(tpy-(CH$_2$)$_5$-SH)$_2$]$^{2+}$ and [Co(tpy-SH)$_2$]$^{2+}$), both of which consisted of a Co-containing coordination complex between two terpyridinyl linker molecules with thiol end groups.\cite{JPark2002}  Using these molecules they were able to observe examples of Coulomb blockade and the single impurity Kondo effect.  Liang \textit{et al.} fabricated a single molecule breakjunction device with individual divanadium (V$_2$) molecules ([(N,N$^{\prime}$,N$^{\prime\prime}$-trimethyl-1,4,7-triazacyclononane)$_2$-V$_2$(CN)$_4$($\mu$-C$_4$N$_4$)]),\cite{Liang2002} and were able to demonstrate charge state transitions by gate modulation, exhibiting the evolution of transport between the Coulomb blockade regime and the Kondo regime.  Since 2002 Kondo resonances have been observed in molecular devices involving a wide variety of molecular active elements, as reported in the literature.\\
\\
\noindent
\textbf{SELECTIVE MANIPULATION OF THE KONDO RESONANCE}\\

Selectively modifying single molecule devices in order to affect changes in the observed zero bias resonance can serve as a means of probing Kondo physics in these systems.  Experimentalists have been able to manipulate both the molecular active element as well as its local environment through their choice of materials and by more direct physical probes.  The resulting impact to the resonance peak has allowed for the investigation of the manner in which Kondo correlations are influenced by these factors in addition to associated phenomena.

\begin{figure}[!t]
\centering
\includegraphics [scale=.34]{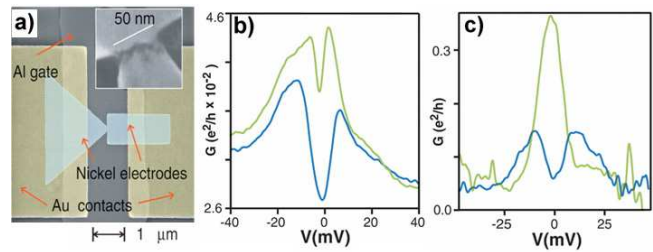}\\
\caption{ \textbf{a)} False color SEM image of break junction fabricated with Ni electrodes.  Inset:  High resolution image after electromigration.  \textbf{b)} and \textbf{c)} Differential conductance \emph{vs.} source-drain bias for two C$_{60}$-based devices with Ni electrodes at $T$ = 1.5 K.  Blue (lower amplitude) lines: electrode magnetizations parallel, \textbf{(b)} $B$ = -310 mT, \textbf{(c)} $B$ = -250 mT.  Green (higher amplitude) lines: electrode magnetizations approximately antiparallel, \textbf{(b)} $B$ = -10 mT, \textbf{(c)} $B$ = -15 mT.  (Reproduced from Ref. 56 with permission from the AAAS)}
\label{NiSplitPeak}
\end{figure}

The zero bias resonance peak is expected to split by approximately $2g{\mu}_BB$ in the presence of an applied magnetic field, as stated previously; however, competing magnetic exchange interactions can significantly alter this response, as demonstrated in an experiment by Pasupathy \textit{et al.}  C$_{60}$-based single molecule breakjunction devices were fabricated using ferromagnetic nickel electrodes (Fig. \textbf{\ref{NiSplitPeak}a}).\cite{Pasupathy2004}  The source and drain electrodes were designed to have different magnetic anisotropies, in order to ensure that they undergo magnetic reversal at different values of $B$.  Although the Kondo effect can be suppressed due to a competing ferromagnetic ground state, zero bias resonances were still achieved in some of these devices in which the molecule is strongly coupled to the electrodes, but the resonances were often found with peak splittings too large to be accounted for by a local field produced by the Ni electrodes.

It was found that the interaction between the spin polarized electrodes and the molecule results in a large local exchange field ($>$ 50~T).  The spin asymmetry in the coupling between the molecule and electrodes, due to quantum charge fluctuations, leads to a spin-dependent renormalization of the molecular energy levels, breaking the spin degeneracy, and results in a splitting of the resonance peak.  Furthermore, the splitting of the zero bias resonance is strongly dependent upon the relative orientation of the electrode magnetizations, which can be controlled with a small external $B$-field ($<$ 100 mT) (Fig. \textbf{\ref{NiSplitPeak}b,c}).  The manipulation of electrode magnetizations is confirmed by measuring the junction magnetoresistance (JMR) of molecule-free electromigrated Ni junctions.  For transport in the Kondo resonance, the JMR can be amplified by turning the electrode magnetization from parallel to antiparallel, since the Kondo resonance occurs closer to the Fermi energy when the magnetization of the leads is antiparallel.  In this respect, the Kondo resonance serves as an indicator when probing the relative orientation of the source and drain electrode magnetizations and their interaction with the local magnetic moment of the molecule.

The Kondo effect can also be used to probe the interaction between the unpaired spin of the local moment and magnetic impurities in the adjacent leads.  The Ruderman-Kittel-Kasuya-Yoshida (RKKY) interaction describes the spin-spin interaction by which magnetic impurities embedded in an electron sea can interact with each other.\cite{Craig2004}  Elastic spin-flip processes leading to the Kondo resonance may be suppressed by competition from the RKKY exchange interaction, which may essentially freeze the spin on the impurity, thereby reducing the conductance near $V_{sd}$ = 0 V and leading to an observed peak splitting.

An investigation by Heersche \textit{et al.} observed Kondo resonances in gold break junction devices resulting from an unpaired spin on a small gold grain formed during the electromigration procedure.\cite{Heersche2006a}  A small number of cobalt impurities were intentionally deposited in the electrodes during the evaporation step of the fabrication process.  They concluded that the (RKKY) interaction between the fixed magnetic impurities in the electrodes and an unpaired spin on the gold grain, which serves as the active element, cause the zero bias Kondo resonance to split.

\begin{figure}[!t]
\centering
\includegraphics [scale=.48]{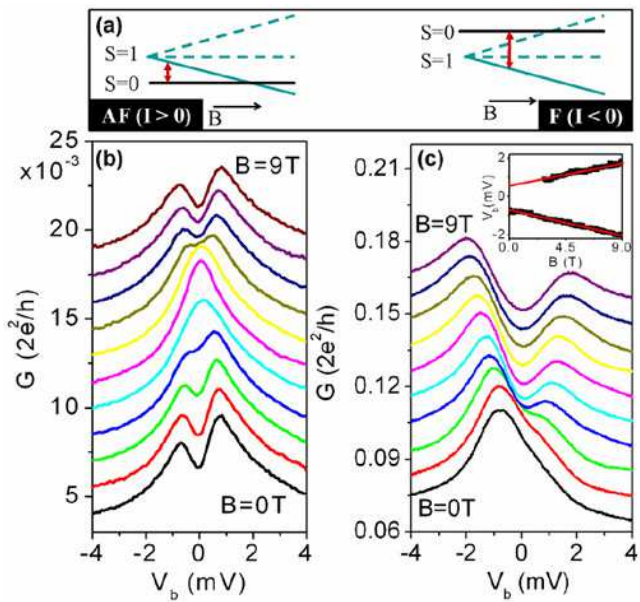}\\
\caption{Magnetic-field dependence of the Kondo resonance in a nanowire break junction with magnetic impurities embedded in the leads.  \textbf{a)} The singlet-triplet transition energy (vertical arrow) decreases, then increases with $B$-field for an antiferromagnetic interaction.  The triplet-singlet energy always increases with field for a ferromagnetic interaction. \textbf{b), c)} Traces of conductance \emph{vs.} applied bias obtained at different values of the external magnetic field, increasing from $B$ = 0~T (bottom line) to $B$ = 9~T (top line) in steps of 0.9~T.  Data taken at $T$ = 250~mK (b) Resonance peak restored at finite field for antiferromagnetic interaction between two spins.  Traces offset for clarity. (c) The peak separation increases linearly with $|B|$ for a ferromagnetic interaction.  Inset: The peak separation at $B$ = 0~T is determined by extrapolating from peak positions to zero field and yields 1.4 $\pm$ 0.3~meV.  (Reproduced from Ref. 57 with permission from the American Physical Society).}
\label{CoSplitPeak}
\end{figure}

The magnetic-field dependence of the spacing between the split peaks depends on the sign of the associated RKKY interaction, $\tilde{I}$.  The interaction can be either ferromagnetic ($\tilde{I} < 0$) or antiferromagnetic ($\tilde{I} > 0$), however both cases will generally suppress the $s$ = 1/2 Kondo effect.  Depending on the type of interaction, the net spin of the active element and the impurity spins together can form a triplet ($S$ = 1) state or an underscreened spin-1 state.  Because the energy between the singlet ground state and the triplet state decreases with $|B|$, an external field can restore the Kondo effect if the RKKY interaction is antiferromagnetic.\cite{Vavilov2005,Simon2005}  The zero bias resonance peak is restored at $B$ = $\tilde{I}/g\mu_BB$, where the singlet and triplet states are degenerate and the external field compensates the antiferromagnetic interaction.  In the case of a ferromagnetic interaction the peak splitting will increase monotonically with $|B|$ because the energy between the triplet and singlet states also increases (Fig. \textbf{\ref{CoSplitPeak}}).

The magnetic field dependence can therefore be used as a probe to distinguish between ferromagnetic and antiferromagnetic interactions between the local moment and conduction electrons.  While the authors have used an entirely metallic system, the local moment could be equally well represented in future experiments by an unpaired spin on a molecule.  The break junction system used in this experiment represents a flexible platform to study the interaction between the spin of the local moment and adjacent static magnetic impurities.

Controlled physical alterations in the structure of single molecule devices provides another means of probing Kondo correlations.  This has been accomplished by both mechanical and electrical means capable of changing the precise arrangement of the source and drain electrodes, as well as the detailed structure and conformation of the active element, and its interaction with the local environment.  Studies of this nature have been performed predominantly with the STM configuration, however the use of a mechanically controlled break junction has also been employed.

Single molecule devices constructed with a STM may not present a scalable device concept; however, they do provide a unique platform for probing the interaction of conduction electrons with molecular orbitals and individual magnetic ions.  A STM possesses the unique ability to both image the DOS of individual molecules on a conducting surface, and to manipulate the detailed chemical environment \emph{via}
mechanical and electrical means.  When the metallic tip is positioned over a magnetic ion or molecule, Kondo exchange interactions may result in a resonance at the equilibrium Fermi level observed in measurements of $dI/dV$.  The characteristic energy scale, $T_K$, extracted from the measured Kondo resonance indicates the relative strength of the coupling between the local magnetic moment and the delocalized conduction electrons.  Repeating these measurements at many points will provide a spatial map of the Kondo resonance, therefore revealing an image of this coupling strength over the surface of molecules.

Several studies of Kondo correlations have exploited the ability of an STM to exert control of the precise surface chemistry and conformation of a single molecule or small group of molecules.  These studies have focused on Co-containing molecules adsorbed on nonmagnetic conducting substrates.  The authors of Zhao \textit{et al}. used a STM to measure differential conductance through a cobalt phthalocyanine molecule adsorbed on a Au(111) surface.\cite{Zhao2005}  Initial measurements exhibited no zero bias resonance since the conduction electrons did not interact strongly with the Co ion.  Dehydrogenation of the ligands (\textit{i.e.} cutting off the hydrogen atoms with voltage pulses from the STM tip) allows the four outward extending orbitals of this molecule to chemically bond to the substrate.  This bonding changes the conformation of the molecule and its coupling to the substrate, and hence its interaction with the conduction electrons.  After this conformational change, a clear Kondo resonance can be observed near the Fermi surface.  Similar work was performed by Wahl \textit{et al}. in which measurements of Kondo coupling in cobalt tetracarbonyl were analyzed before and after tip-induced dissociation of the molecule.\cite{Wahl2005}

\begin{figure}
\centering
\includegraphics [scale=.295]{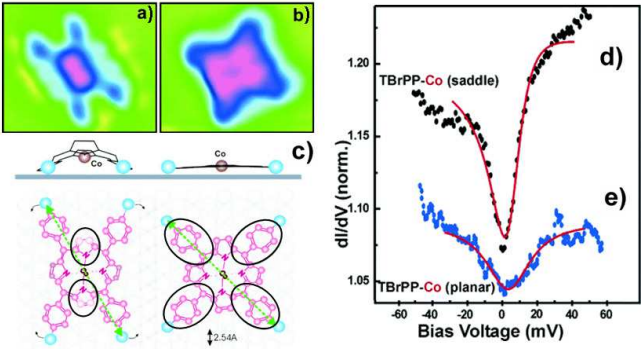}\\
\caption{\textbf{a)} STM images of saddle conformation (width 11 \AA, length 18 \AA) and \textbf{b)} planar conformation (length 15.5 \AA) of TBrPP-Co on Cu(111). \textbf{c)} Corresponding models of saddle (left) and planar (right) conformation.  Blue and pink balls represent bromine and carbon atoms, respectively. The black ovals indicate the areas providing higher current in the STM images.  The length of the molecule along the diagonal (dashed green arrow) is unchanged upon switching between the two conformations.  Traces of $dI/dV$ \emph{vs.} applied bias exhibiting Kondo signatures for \textbf{d)} a saddle TBrPP-Co molecule and \textbf{e)} a planar TBrPP-Co molecule.  The solid lines represents the Fano line-shape fits to the data.  The spectra for the saddle and planar conformations have been vertically offset by 0.07, 0.04, respectively.  Measurements were taken at the center of each molecule.  (Reproduced from Ref. 63 with permission from the American Chemical Society)}
\label{Conformation}
\end{figure}

A related experiment investigated the TBrPP-Co (5,10,15,20-tetrakis(4-bromophenyl)porphyrin-Co) molecule, which is
found in two conformations, ``planar'' and ``saddle'', both of which anchor themselves onto a Cu(111) surface \emph{via} their four bromine atoms (Fig. \textbf{\ref{Conformation}a, b,} and \textbf{c}).  A single voltage pulse of 2.2~V from a STM tip positioned above one of these molecules has the ability to switch the complex between the two conformations.\cite{Iancu2006a}  A Kondo resonance can be observed in molecular devices made with both conformations, but they possess distinctively different values of $T_K$, and hence experience different interactions with the substrate.  The Kondo temperature is decreased in the case of the saddle conformation as compared to the planar TBrPPCo (Fig. \textbf{\ref{Conformation}d, e}).

The authors suggest that the measured Kondo resonances in these molecules differ because of the potential for conduction to originate from more than one path.  They indicate that a resonance may be generated by spin-electron coupling to the substrate through bonding orbitals (\emph{via} the bromine atoms) and by direct coupling to the Co ion.  When in the saddle conformation, the Co atom is lifted away from the surface as the porphyrin plane is bent upward, reducing the coupling between the Co atom and the free electrons of the Cu(111) substrate (Fig. \textbf{\ref{Conformation}c}).  Coupling \emph{via} molecular orbital bonding is therefore likely to be the only conduction pathway available in this state.  For the planar (higher $T_K$) conformation the Kondo effect may result from the orbital bonding as well as Co-surface interaction.  This single molecule switching mechanism reveals a means of altering the relevant interactions without destroying or changing the chemical composition.

A similar study indicated that the chemical environment \textit{near} a molecule also plays an important role with respect to Kondo correlations.  Beginning with a 2D molecular assembly of TBrPP-Co,
\begin{figure}[b]
\centering
\includegraphics [scale=.42]{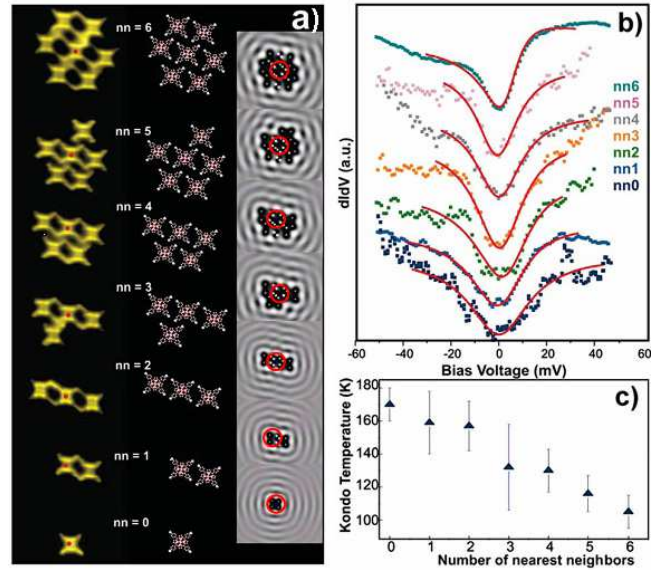}\\
\caption{\textbf{a)} Sequence of STM images (left), created by removing one nn molecule at a time with the STM tip, and corresponding models (middle).  The calculated electron standing wave patterns (right) reveal an increasing exposure of the center molecule (red circles) to the surface state electrons.  White and black colors in the calculated images represent higher and lower electron densities, respectively.  \textbf{b)} The $dI/dV$ spectra are measured at each step by positioning the tip above the center molecule (red dots in \textbf{(a)}).  The spectra are offset for clarity.  Horizontal displacements of 3 to 10 meV are taken for the spectra representing nn6, nn5, nn4, nn2, and nn1.  (c) $T_K$ plotted as a function of the number of nearest neighbors.  (Reproduced from Ref. 64 with permission from the American Physical Society)}
\label{2Dassembly}
\end{figure}
where the center molecule is initially surrounded by six adjacent molecules,\cite{Iancu2006b} Iancu \textit{et al.} use a STM tip to measure the Kondo resonance, and thus $T_K$, above the center molecule.  One nearest neighbor (nn) molecule at a time is removed from the assembly, after which $T_K$ is again measured above the molecule originally at the center (Fig. \textbf{\ref{2Dassembly}}).  Measurements of $T_K$ before decomposition of the assembly indicate values that are the same as those measured inside a full ribbon structure.

As the number of nn molecules is reduced, $T_K$ is increased in a controlled manner from 105~K to 170~K.  It is found that manipulating the nearest neighbor configuration has a direct effect on the coupling strength between the local moment and conduction electrons responsible for the Kondo resonance.  The molecule located inside the cluster will have a reduced interaction with the delocalized surface state electrons as compared to the molecules around the edges.  This again represents a means of affecting the interactions responsible for the Kondo effect without affecting the chemical composition.

\begin{figure}[t]
\centering
\includegraphics [scale=.4]{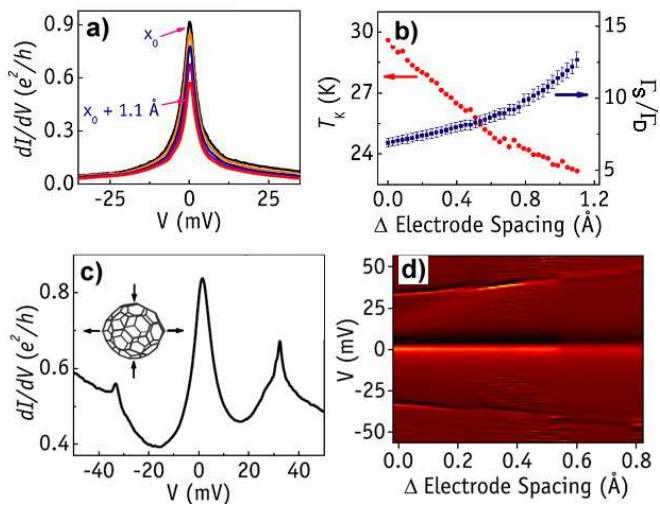}\\
\caption{\textbf{a)} $dI/dV$ traces at various electrode spacings and $T$ = 1.6 K for a single mechanically controlled break junction device.  Smallest spacing corresponds to tallest peak.  Largest spacing corresponds to shortest peak.  \textbf{b)} The Kondo temperature, $T_K$, and the relative coupling, $\Gamma_M$/$\Gamma_L$, as a function of electrode spacing for same device at $T$ = 1.6 K.  The uncertainty in the determination of $T_K$ values is $<$ 0.4 K.  \textbf{c)} $dI/dV$ for another device at $T$ = 1.6 K, exhibiting satellite peaks near $\pm$33 mV.  Left inset: Schematic of intracage vibrational mode.  \textbf{d)} $d^2I/dV^2$ as a function of bias voltage and electrode spacing for device in \textbf{(c)}. (Reproduced from Ref. 6 with permission from the American Physical Society).}
\label{MechKondo}
\end{figure}

A mechanically controlled break-junction can also selectively manipulate the molecule-electrode coupling.  The controlled movement of a piezo actuator is used to stretch a nanowire until a small gap appears.  In the case of J. J. Parks \textit{et al}.\cite{Parks2007} this technique was employed to produce a
C$_{60}$-based single molecule device exhibiting a spin-1/2 Kondo resonance, after which the interelectrode gap size was further modified (with picometer precision) to change the geometry of the molecule-electrode assembly.  Initial values of $T_K$ are extracted from measurement of the equilibrium conductance as a function of temperature.  Increasing the spacing between source and drain
electrodes is shown to tune both the height and width of the Kondo resonance, effectively decreasing $T_K$ by as much as 37\% in one example.  The mechanical manipulation thus has a direct impact on the Kondo exchange correlations by altering the relative coupling strength between the molecule and the two electrodes.  A standard theoretical model is able to characterize the changes in molecular orbital-electrode interaction (and thus $T_K$) due to electrode
displacement.

Excited states, manifested as finite-bias peaks in $dI/dV$, appeared in some devices at energies consistent with the 33-35~mV intracage vibration for C$_{60}$, as in Refs.~8 and 11.  Similar vibrational sidebands have recently been observed in STM measurements of Kondo resonances in organic charge-transfer assemblies.\cite{FT2008}  The mechanical motion also affects the energy of these active vibrational modes, but not in accord with predictions based on a simple semiempirical Hamiltonian.  This suggests that basic approaches used to capture transport properties in these devices may be insufficient, and that new models or approaches will ultimately be required to provide a complete explanation.  Further observations of transport in the Kondo regime of single molecule devices, as discussed in the following section, have led to similar conclusions.\\
\\
\noindent
\textbf{OBSERVATIONS AND COMPARISONS TO STANDARD MODELS}\\

The basic theoretical descriptions of the Kondo state, outlined above, have enjoyed a good deal of success in their agreement with measured transport properties in experiments using semiconductor QDs.  This, together with the advantage of tunable tunneling barriers and a high level of reproducibility, has made electrostatically defined semiconductor QDs a canonical system for studying the Kondo state.  Transport measurements in QDs made in materials like GaAs have provided a strong foundation for researchers investigating analogous effects in molecular devices.  Experimental investigations of Kondo correlations in devices where a single molecule is utilized as a type of ``natural'' QD reveal transport properties for which the models described earlier cannot fully account.   These results indicate that sophisticated analysis techniques together with a more thorough understanding of the underlying physics are required to resolve these molecular complications.

\begin{figure*}
\centering
\includegraphics [scale=.44]{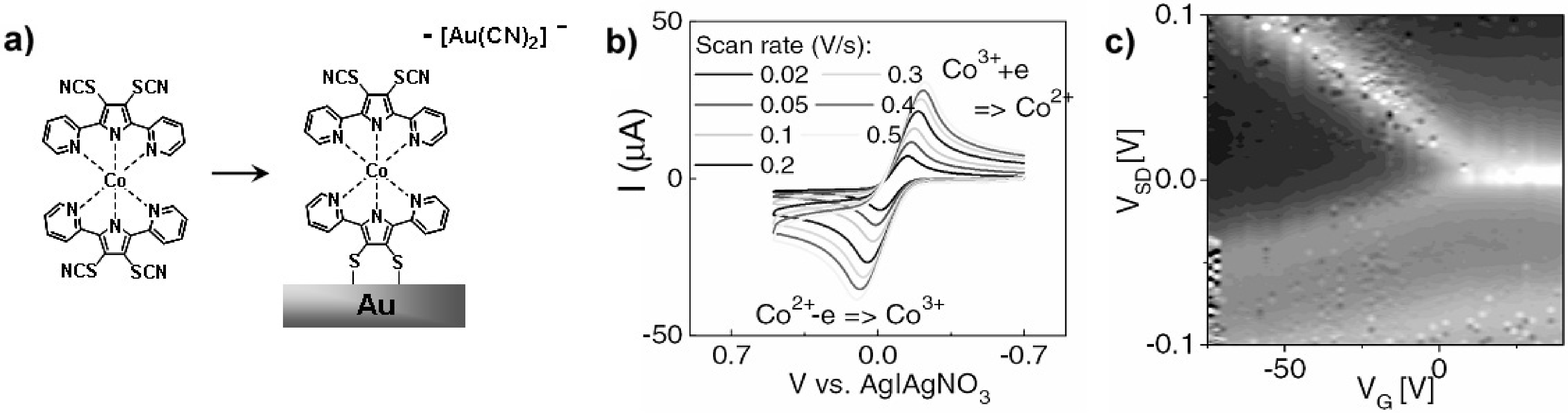}\\
\caption{\textbf{a)} Assembly sequence on gold and final assembled structure of the Co-containing complex.  (From Ref. 14) \textbf{b)} Cyclic voltammograms showing reversible change in Co ion charge state.  \textbf{c)} Stability diagram of a device containing a single complex exhibiting a gate induced transition between the Coulomb blockade regime and the Kondo regime ($T$ = 5 K).  White corresponds to $dI/dV$ = 3 $\times$ 10$^{-6}$ S.  The zero bias peak in the right-hand charge state disappears at the charge degeneracy point.  (From Ref. 71).}
\label{TransportMol}
\end{figure*}

One discrepancy between physical descriptions sufficient for semiconductor QDs and SMTs can be seen in the functional form of the low energy nonequilibrium conduction.  Scaling can be seen in the response of the Kondo system to different perturbations.  Applied bias, temperature, and magnetic field may perturb the low energy conductance of a Kondo resonance in qualitatively different ways, but the exponents of their lowest order responses are identical.  Theoretical treatments suggest that the low energy region of every Kondo system can be described by the same universal scaling function (\emph{i.e.} conductance collapses onto a single curve) when it is normalized by the characteristic energy scale, $k_BT_K$\cite{Ralph1994,Schiller1995}  Limited experimental work with GaAs QDs verify that the non-equilibrium conductance of the single channel spin-1/2 Kondo resonance at low bias and temperature adheres to a scaling function characterized by two parameters.\cite{Grobis2008}

We recently tested the ability of a single scaling function to describe transport in the spin-1/2 Kondo regime of single molecule devices.  Differential conductance was analyzed as a function of both temperature and applied bias for a large number of samples (29) fabricated on a Si/SiO$_2$ substrate.  The electromigration technique was employed to create nanometer-size gaps in Au nanowires while alternately using two different molecular active elements: C$_{60}$ and a Cu-containing version of the transition metal complex in Fig. \textbf{\ref{TransportMol}a} (bis(2,5-di-[2]pyridyl-3,4-dithiocyanto-pyrrolate)Cu(II)).  The C$_{60}$ molecules can be strongly bound to the Au electrodes by charge transfer from the metal surface to the molecule.\cite{Altman1993}  The conjugated complexes self-assemble on the Au nanowires in tetrahydrofuran (THF) through loss of the $^-$CN moieties and formation of Au-S covalent bonds (Fig. \textbf{\ref{TransportMol}a}).\cite{Ciszek2004,Yu2004c}  These devices were found to have a variety of Kondo temperatures, ranging from 35~K to 155 K, based on the temperature-dependence of their zero bias conductance.  They were also found to possess a range of asymmetries, as inferred from the magnitude of the resonance peak as $T \rightarrow$ 0 K, according to Eq. (\ref{Asymmetry}).\cite{Beenakker1991}

\begin{figure}[b]
\centering
\includegraphics [scale=.172]{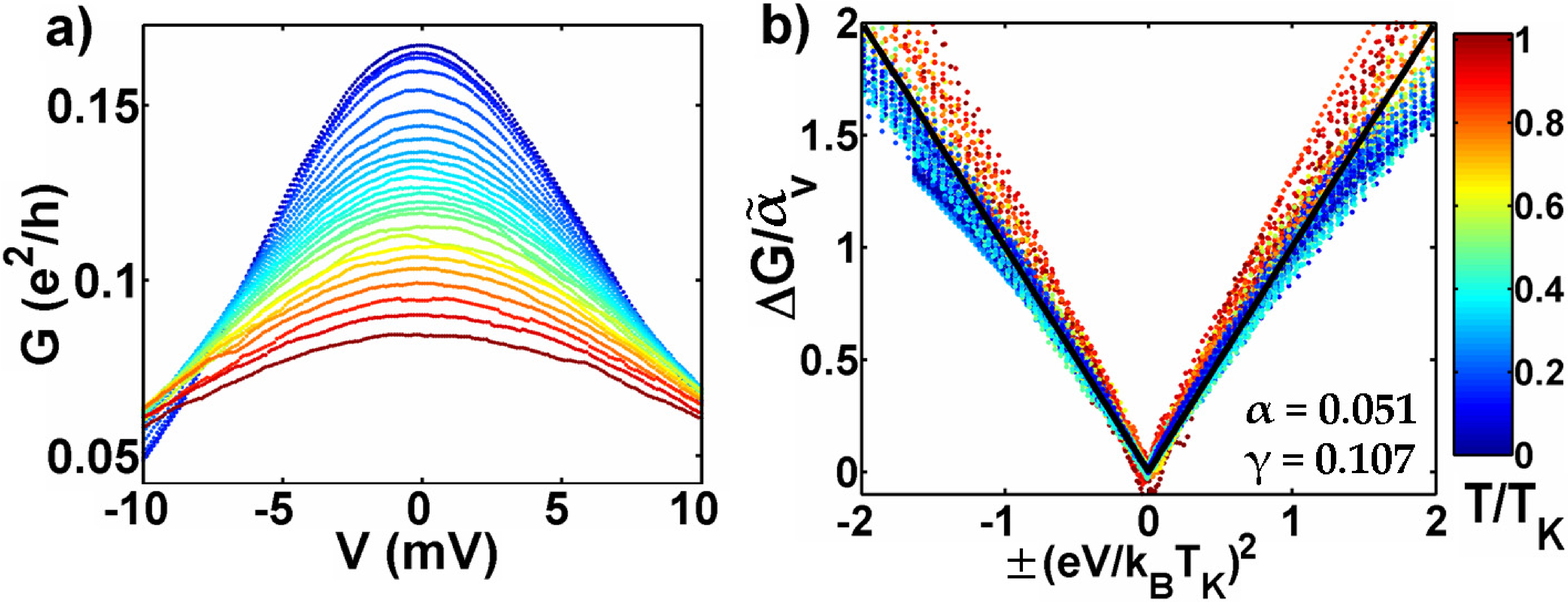}
\caption{\textbf{a)} Conductance as a function of applied bias for $T/T_{\mathrm K} \lesssim 1.0$.  A background contribution to the conductance, $G_{\mathrm b}$, has been subtracted off.  \textbf{b)} Scaled conductance, $\Delta{G}/{\tilde{\alpha}_V}$ \emph{vs.} $(eV/k_{\mathrm B}T_{\mathrm K})^2$ for four samples, including the data from device in (a), using the average extracted values of the scaling coefficients, where $\Delta{G} = (1-G(T,V)/G(T,0))$.  The solid black line represents the associated universal curve.  The color bar in (b) pertains to both figures.  (Adapted from ref. 50).}
\label{Scaling}
\end{figure}

Transport measurements of the temperature and bias dependence of the non-equilibrium conductance are in agreement with the functional form of the expected scaling function.  Namely, a quadratic power law in temperature and bias characterized by two scaling coefficients.  However, values of the extracted scaling coefficients are found which are consistent throughout the data set, but differ systematically from both theoretically predicted values and coefficient values extracted from measurements of nonequilibrium transport through a GaAs dot.\cite{Grobis2008,Scott2009}  The low energy approximation of the scaling function can be written as \cite{Nagaoka2002,Scott2009,Schiller1995,Grobis2008,Majumdar1998}
\begin{eqnarray}
\frac{G(T,0) - G(T,V)}{c_TG_{\mathrm 0}}~\approx ~~~~~~~~~~~~~~~~~~~~~~~~~~~~~~~~~~~~~ \nonumber \\
\alpha\left(\frac{eV}{k_{\mathrm B}T_{\mathrm K}}\right)^2 - c_T\gamma\left(\frac{T}{T_{\mathrm K}}\right)^2\left(\frac{eV}{k_{\mathrm B}T_{\mathrm K}}\right)^2.
\label{GTV2}
\end{eqnarray}
where $\alpha$ and $\gamma$ are the scaling coefficients and $c_T \approx 4.92$ is a constant.  We calculated average values of $\alpha = 0.051 \pm 0.01$ and $\gamma = 0.107 \pm 0.027$.  This is in contrast to values of $\alpha = 0.10 \pm 0.015$ and $\gamma = 0.5 \pm 0.1$, for a GaAs QD.  The consistently smaller values of $\alpha$ and $\gamma$ found for all of these molecular devices, compared to the GaAs case as well as predictions based on the Anderson\cite{Costi1994,Konik2002,Oguri2005} and Kondo\cite{Schiller1995,Majumdar1998,Doyon2006,Pustilnik2004} models, implies that for a given $T_K$ inferred from the temperature dependence of the equilibrium conductance, the resonance peaks observed here were broader in voltage, and evolved more slowly with temperature.

The uniformity of extracted coefficient values for SMTs made with the two different molecules suggests that the origins of this discrepancy likely do not depend on the detailed molecular structure of the active element.  The size of the on-site repulsion, $U$, of the molecule relative to the single-particle level spacing, $\Delta$, may play a role.  Generic vibrational modes in SMTs, such as a center-of-mass oscillation, may also have an effect by modifying the temperature dependence leading to a different effective $T_K$ than is found in the bare Kondo system, as discussed below.

Further experiments using the same transition metal complex as above, alternating between versions of the complex containing a Cu and a Co ion, serve to highlight some of the distinguishing transport characteristics of the single molecule system.  Again, single molecule breakjunction devices are fabricated on a Si/SiO$_2$ substrate using the electromigration technique.  Zero bias resonance peaks were observed to have a temperature and magnetic field dependence consistent with the spin-1/2 single impurity Kondo effect.  The start (or end) of the resonance, with respect to a gate bias, was always coincident with the charge degeneracy point (Fig. \textbf{\ref{TransportMol}c}), and similar features were not seen in any control devices.

Evolution of the Kondo resonance with gate bias was found to be inconsistent with the simple model, which works well for analogous data obtained in semiconductor QDs and carbon nanotube devices.  More specifically, it was found that $T_K$($V_g$) was distinctly weaker than that expected according to Eq. (\ref{Haldane}).\cite{Yu2005}  It has been suggested that the specific orientation of the active element relative to the detailed electrode geometry can result in poor gate coupling prohibiting effective gate modulation.\cite{Perrine2007}  Anomalously weak gate dependence may also be traced back to a number of potentially incorrect assumptions made when invoking Eqn.~(\ref{Haldane}).  An invalid value of $\varepsilon_0/\Gamma$, which is inferred by normalizing $V_g$ by the width of the Coulomb blockade charge degeneracy point, may be due to some mechanism intrinsic to the SMT system.  Screening correlations in the mixed valence regime can renormalize the measured $\Gamma$ to a value different than the $\Gamma$ relevant to the Kondo temperature.

The energy of intramolecular exchanges may be another factor requiring further consideration.  This energy is typically ignored, yet may be comparable to $U$ and $\Delta$ in single molecule devices.  Electronic structure calculations confirm that the coupling to intramolecular excitations is strong in both the Co and Cu-containing complexes, complicating the resulting molecular orbital degeneracies and/or splittings.  Ligand field effects, for example, may split the degeneracy of the $d$ states, and delocalization of the majority spin states, extending into the ligands, is consistent with large $\Gamma$ values for these systems.  Degeneracies for the unpaired spin in the molecular orbitals hybridized with the metallic electrodes may enhance $T_K$, making the normalization of the abscissa correspondingly too large.\cite{Bonca1993}  Intramolecular exchange and the coupling of vibrational modes to charge may both increase $T_K$ and decrease its gate dependence compared to an identical system with no vibrational coupling.\cite{Cornaglia2004,Cornaglia2005}  Among the possible influences discussed here, a considerable amount of attention has been paid to the impact of vibrational effects.

Local vibrational modes are a distinctive feature in the single molecule system distinguishing it from the otherwise analogous case of the semiconductor quantum dot.  In semiconductor dots, optical phonon modes are considerably higher in energy than the Coulomb charging scale, and are therefore typically ignored in theoretical models of the low energy dot properties.  In the molecular case, however, local vibrational modes can have energies $\sim$tens of meV, often considerably lower than the Coulomb scale.  Molecules can move and distort upon the addition or removal of an electron, leading to phonon assisted tunneling, which is commonly observed in stability diagrams plotting conductance in the Coulomb blockade regime.  Conductance features related to intramolecular vibrations\cite{Parks2007,Yu2004c} have also shown up in the Kondo regime, and similar excitations expose other unresolved transport phenomena.  

\begin{figure}[t]
\centering
\includegraphics [scale=.152]{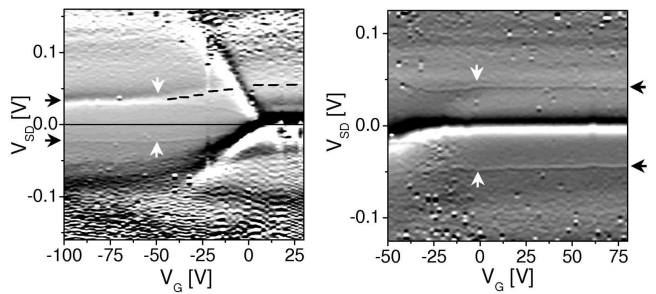}\\
\caption{Colormaps of $d^2I/dV_{sd}^2$ as a function of $V_{sd}$ and $V_g$ at 5~K for two devices.  Smoothing window in  $V_{sd}$ is 5 mV.  Brightness scales are -8 $\times$ 10$^{-5}$ A/V$^2$ (black) to 3 $\times$ 10$^{-5}$ A/V$^2$ (white), and -2 $\times$ 10$^{-5}$ A/V$^2$ (black) to 2 $\times$ 10$^{-5}$ A/V$^2$ (white), respectively.  The zero bias features correspond to Kondo peaks in $dI/dV_{sd}$.  Black arrows indicate prominent inelastic features.  When these inelastic features approach the boundaries of the Coulomb blockade region, the levels shift and alter lineshape (white arrows).  Black dashed line in left map traces an inelastic feature across the Coulomb blockade region boundary and into the Kondo regime.  (From Ref. 71)}
\label{IETS}
\end{figure}

Inelastic electron tunneling spectroscopy (IETS) reveals features in a portion of the SMTs studied, which correspond energetically to vibrational excitations of the molecule, as determined by Raman and infrared spectroscopy.  The tunablity of these systems allows for the examination of such features across successive charge states. Inelastic cotunneling processes can produce peaks, dips, or intermediate structures in $d^2I/dV^2$, which are observed in the blockaded regime of an even charge state and persist into the Kondo regime in the form of satellite features paralleling the zero bias resonance (Fig. \textbf{\ref{BasicTransport}b,c}).\cite{Yu2004c}  These features are a signature of IETS, and can be seen to undergo a clear change in lineshape and intensity after passing the charge degeneracy point (Fig. \textbf{\ref{IETS}}), indicating a complicated conduction process and non-trivial transition between the two conduction regimes.

Recent analytical and numerical investigations of transport in single molecule devices have considered the role of electron-phonon coupling and its relationship with Kondo correlations.  These effects can be modeled with appropriate modifications to the Anderson Hamiltonian.  A term may be added to Eq.~(\ref{Hamiltonian1}) to account for the local vibrational modes of the molecule,
\begin{equation}
H^{}_{v} = \sum_{l}\hbar{\omega^{}_l}{a^\dag_l}{a^{}_l},
\label{Hv}
\end{equation}
which are represented here as a discrete number of harmonic oscillator modes.  The creation and annihilation operators $a^\dag_l$ and $a^{}_l$ can be used to define a displacement operator for a given mode:
\begin{equation}
x^{}_l = \sqrt{\frac{\hbar}{2 m^{}_{l}\omega^{}_{l}}}( a^\dag_l + a^{}_l).
\label{displacement}
\end{equation}
Here $m_{l}$ is an effective mass characterizing the local vibrational mode of frequency $\omega_{l}$.  Another term, utilizing the displacement operator, is employed to describe the coupling between these local modes and the molecular energy levels,
\begin{equation}
H^{}_{ev} = \sum_{lv\sigma}{\lambda^{}_{lv}}x^{}_l\hbar{\omega^{}_l}d^\dag_{v\sigma}d_{v\sigma}. 
\label{Ht}
\end{equation}
Here $\lambda$ is a dimensionless matrix of coupling constants that links molecular level $v$ with local vibrational mode $l$.  

The local displacement of the molecule due to vibrations, including nanomechanical oscillations, can modulate the molecule-electrode tunneling barriers, so the portion of the Hamiltonian that describes the tunneling of electrons between these components may be altered to incorporate an explicit dependence on Eq.~(\ref{displacement}).\cite{Natelson2006b}  Terms are also included to account for bulk vibrational modes (phonons) of the electrodes
\begin{equation}
H^{}_{ph} = \sum_{q}{\hbar}{\Omega^{}_q}b^\dag_lb^{}_l,
\label{Hph}
\end{equation}
and their coupling to local vibrational modes of the molecule,
\begin{equation}
H^{}_{vph} = \sum_{lq}\beta^{}_{lq}x^{}_lX^{}_q.
\label{Hvph}
\end{equation}

In order to make the resulting Hamiltonian tractable, a number of these treatments consider the coupling of vibrations to only a single electronic level, and have utilized a modified version of the Anderson Hamiltonian (the Anderson-Holstein model) that incorporates one vibrational mode of the molecule linearly coupled to charge fluctuations.\cite{Cornaglia2005b,Balseiro2006,Regueiro2007,Silva2009}  In this approximation, summations over the $v$ and $l$ indicies are absent.  These studies confirm that the nature of the Kondo effect can be modified by electron-phonon interactions, and the resulting effects will depend on the symmetry and frequency of the phonon mode and on the strength of the coupling. 

In the regime of weak electron-phonon coupling ($U \gg 2\lambda^2/\hbar\omega_0$) the same basic Kondo effect is expected to occur for a spin-1/2 impurity, but the interaction potentially results in an increase of the spin fluctuation energy and of the amplitude of charge fluctuations.  In the case of equal source-drain coupling, the effective charging energy is reduced to a new value,
\begin{equation}
U^{}_{eff} \approx U - 2{\lambda}^2\hbar{\omega^{}_0}.
\label{Ueffective}
\end{equation}
The calculated effect also leads to an increase in the value of $T_K$ and a renormalization of the electronic energy level, $\varepsilon_0$, and the single-particle level width, $\Gamma$.  These changes have been shown to correlate to an altered $T_K(V_g)$ and a weakening of the influence from an applied gate voltage, as observed in Ref. 76.  Another change in gate dependence may manifest as an asymmetry in the curve of $G(V_g)$ about its maximum.\cite{Cornaglia2005b}

In the strong electron-phonon coupling regime ($U \ll 2\lambda^2/\hbar\omega_0$) the properties of the device are described in terms of a charge analog of the Kondo effect.\cite{Cornaglia2005,Cornaglia2007}  In this regime the transport characteristics of the molecular junction are controlled by the large charge polarizability of the molecule.  The onsite repulsion, $U$, can be renormalized to a negative value.  This means that the effective $U$ produces an attractive interaction between the electrons, which is described as a polaronic effect and thus leads to qualitatively different physics.  One consequence, for example, is a general insensitivity of the conductance to an externally applied magnetic field.\cite{Cornaglia2005}  More sophisticated modeling techniques, such as electronic structure calculations involving molecules bound to realistic electrodes while including complex correlations, would also greatly facilitate this research. 

\begin{figure*}
\centering
\includegraphics [scale=.60]{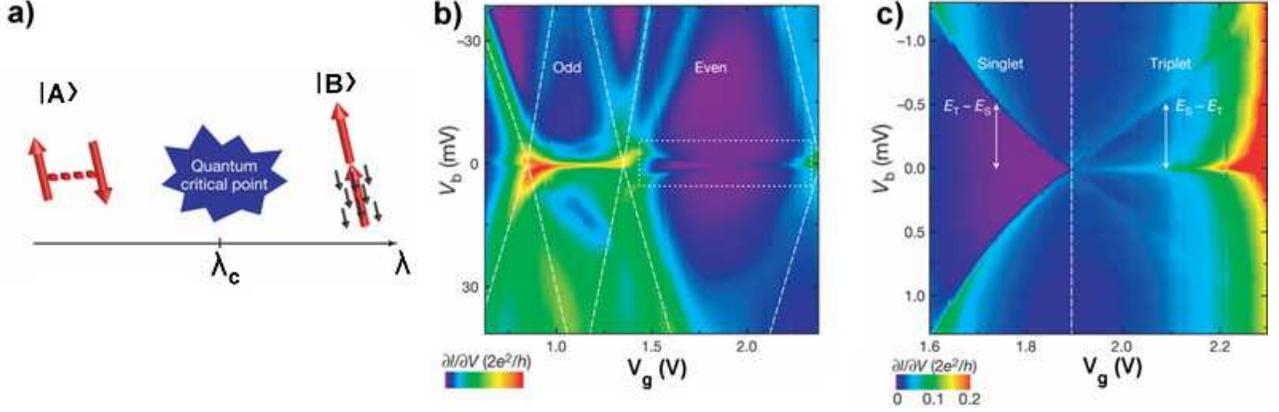}\\
\caption{\textbf{a)} Quantum phase transition depicting quantum state $|A>$ driven by a non-thermal external parameter $\lambda$ to another quantum state $|B>$ with a different symmetry, passing through a critical point at $\lambda = {\lambda}_c$.  In the SMT devices here, $|A>$ is a singlet state and $|B>$ is a triplet state that is partially screened by one conduction electron channel, represented by black arrows.  \textbf{b)} Colormap over two Coulomb diamonds plotting $dI/dV$ (in units of 2$e^2/h$) as a function of bias voltage $V$ and gate voltage $V_g$ at $T$ = 35 mK and $B$ = 0 T.  \textbf{c)} Details of the colormap in the dotted white rectangle in (c), showing the singlet to triplet spin transition.  (Reproduced from Ref. 97 with permission from the Nature Publishing Company).}
\label{Balestro1}
\end{figure*}

Further experimental data will be needed to provide support to existing theories and insights into the distinctive nature of transport in molecular electronic devices.  For example, an analysis of shot noise in molecular devices would enable a new method of studying exchange correlations, and may serve to contrast fundamental differences in transport behavior between single molecule devices and semiconductor QDs.  Shot noise is the result of current fluctuations that occur due to the discrete nature of charge carriers, and can provide a direct measurement of the effective charge ($e^\ast$) in a system.   For a completely uncorrelated stochastic process, current is characterized by Poissonian shot noise.  However, electron-electron interactions can correlate the charge transport process and suppress shot noise.

In Djukic \textit{et al.}, shot noise was measured in single molecule devices consisting of a D$_2$ molecule between Pt source and drain electrodes.\cite{Djukic2006}  Their measurements indicate that the shot noise is strongly suppressed with respect to the classical noise levels, and that transport through the Pt-D$_2$-Pt system is carried predominantly by one nearly transparent conduction channel.  Despite its Fermi liquid nature, many body effects, such as Coulomb repulsion and the Pauli exclusion principle, persist at low energies in the Kondo regime, and the effective charge is expected to be 5/3$e$ due to these interactions.\cite{Sela2006}  This is predicted to be a universal result in the Kondo system, independent of $T_K$.  Measurements of the shot noise in the Kondo regime of a single molecule device would provide a test of this theory and contribute to the study of correlations induced in mesoscopic transport by different types of interactions.\\
\\
\noindent
\textbf{BEYOND THE SINGLE CHANNEL SPIN-$\frac{1}{2}$ SYSTEM}\\

We have heretofore considered primarily experimental results concerning the single channel spin-1/2 Kondo effect, as described by a localized Anderson impurity screened by delocalized conduction electrons.  But the Kondo effect can occur when there is more than a single channel coupled to the impurity, and it can be observed due to the screening of a localized impurity with spin state $S > 1/2$.  Moreover, it can result from non-spin degrees of freedom, such as the SU(4) Kondo effect due to orbital degeneracies in carbon nanotube devices.\cite{Makarovski2007}  Until recently, experimental observations of these more exotic Kondo systems have been confined to interactions with semiconductor QDs and carbon nanotubes.\cite{Potok2007,Makarovski2007}  This can be attributed, in part, to the limited ability with which one can tune the coupling strength between the active element and the source, drain, and gate electrodes in single molecule devices.  The relatively high energy scales at which most SMT experiments have been conducted has also made the properties associated with these variant systems difficult to observe.

This review has focused on the relatively common case of a single screening channel coupled to a single spin-1/2 impurity.  As $T \rightarrow 0 K$ this describes what may be called a fully screened Kondo effect.  More generally, if the number of screening channels coupled to an impurity state is $n$ and the impurity (molecule) is in spin state $S$, the case of $2S = n$ corresponds to a completely screened Kondo effect.  When $n < 2S$ the impurity is said to be underscreened.  Likewise, when $n > 2S$ there are an excess number of screening channels and the impurity is overscreened.  The transport properties in the Kondo regime, namely the conductance measured as a function of temperature and bias, will depend on the spin state of the impurity and on the number of screening channels utilized by the conduction electrons.  Recent reports have observed single molecule devices in even charge states exhibiting Kondo physics.

Both Refs. 91 and 92 each involve devices containing a single C$_{60}$ molecule with $S = 1$, coupled to conduction electrons.  This total spin arises because of the occupation of two different molecular orbitals each with one electron, with Hund's rule favoring the triplet ground state of total spin 1.  Each orbital can have different couplings to the conduction electrons in the source and drain, leading to two different energy scales, $T_{K1}$ and $T_{K2}$.\cite{Wiel2002,Grobis2007} These energy scales depend exponentially on coupling strength, so that it is not difficult experimentally to be in the situation where $T_{K1} \ll T_{K2}$.  If the experiments probe temperatures $T_{K1} \ll T < T_{K2}$, only one screening process is apparent, and the zero-bias resonance measured in Refs. 91 and 92 corresponds to an underscreened spin-1 Kondo effect.

\begin{figure}
\centering
\includegraphics [scale=.28]{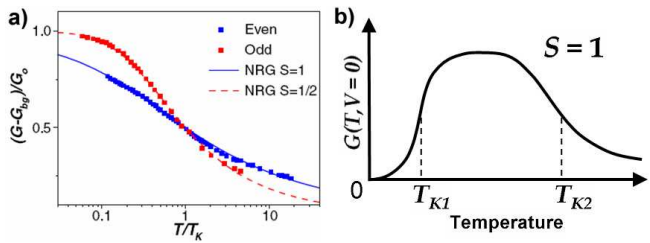}\\
\caption{\textbf{a)} Equilibrium conductance \emph{vs.} temperature measured in
the odd (red squares) and even (blue squares) charge states of sample A in Ref. 91.  Solid and dashed lines are fits to the numerical renormalization group analysis for a fully screened $S = 1/2$ impurity and an underscreened $S = 1$ impurity, respectively.  See reference for details.  \textbf{b)} Two screening channels with different couplings to the local moment, and thus different $T_K$, can allow the conduction electrons to fully screen an $S = 1$ impurity for $T < T_{K1}$, leading to a zero-bias conductance that changes non-monotonically with temperature. (Reproduced from Refs. 91 and 95 with permission from the American Physical Society).}
\label{HigherSpin}
\end{figure}

The functional form of the equilibrium conductance as a function of temperature can be distinguished from the form used to describe the fully screened spin-1/2 effect by the more gradual slope and a value of $G(T,0)$ that does not saturate as $T$ goes to 0 K (Fig. \textbf{\ref{HigherSpin}a}).  If temperatures below $T_{K1}$ can be reached then the second screening channel may allow for the impurity to become fully screened as $T \rightarrow 0$.  A fully screened spin-1 impurity produces no resonance, so the curve of $G(T,0)$ will no longer increase monotonically with decreasing $T$.\cite{Pustilnik2001}  This can be described as a two-stage Kondo effect (Fig. \textbf{\ref{HigherSpin}b}).

To date, few experiments with single molecule break junction devices have been carried out in the millikelvin temperature range,\cite{JPark2002} where some of these non-spin-1/2 Kondo effects may become more accessible.  This is due in part to the large number of devices typically required for a successful analysis, and to the fact that the relatively high Kondo temperatures frequently associated with single molecule devices have made experiments in this temperature range an unnecessary complication.  Recently however, in Roch \textit{et al.}\cite{Roch2008b} investigators were able to conduct a detailed analysis of Kondo physics in C$_{60}$-based electromigrated breakjunction devices by performing transport measurements below 100 mK using a dilution refrigerator.  They observed multiple charge state transitions and the clear evolution from a Coulomb blockade in even charge states to a Kondo resonance in odd charge states.  The low energy regime in which their transport measurements are conducted enabled them to observe a number of novel effects.

In one of their experiments, a detailed study of transport in an even charge state of a SMT reveals Kondo exchange interactions, evident in the stability diagram, attributed to higher spin ($S > 1/2$) Kondo states.\cite{Roch2008a}  Most significantly, a quantum phase transition induced by gate voltage tuning is observed between two competing ground states (singlet and triplet electron spin states) involving Kondo correlations.  This represents an observation distinctly different from previously observed Kondo transitions in semiconductor and nanotube quantum dots.\cite{Sasaki2000,Nygard2000}  In the singlet ground state no zero bias resonance is observed at base temperature.  However, when the applied source-drain bias exceeds the energy difference between the singlet ground state and triplet (excited) state, a non-equilibrium Kondo effect is observed, producing $dI/dV$ peaks at finite bias.  This finite-bias Kondo effect has also been reported by Osorio \textit{et al.} in transport measurements through a thiol end-capped oligophenylenevinylene molecule.\cite{Osorio2007}  In the triplet ground state, a zero-bias peak due to Kondo screening is clearly visible, as are finite bias peaks related to the singlet (excited) state when the source-drain bias exceeds the triplet-singlet energy difference.

A theory for a related gate-induced quantum phase transition in a SMT has been been put forth by Kirchner \textit{et al.}, in which a molecule between an antiparallel arrangement of ferromagnetic leads could couple to both the usual
fermionic electron bath as well as the bosonic spin waves in the leads.\cite{Kirchner2005}  The authors propose that transport measurements in the Kondo regime could reveal the crossing of the critical point as a predicted change in the exponent of the fractional power law governing the behavior of the equilibrium Kondo conductance as a function of temperature.\\
\\
\noindent
\textbf{CONCLUSION}\\

Renewed interest in the highly correlated electron state leading to the Kondo effect has sparked a number of studies, both experimental and theoretical, addressing the manybody problem leading to this emergent phenomenon.  First observed in an electrostatically defined GaAs QD, recent advances in nanofabrication techniques have enabled the experimental realization of the single impurity spin-1/2 Kondo effect in molecular devices.  Here we have reviewed experimental advances in the study of the Kondo state as observed in molecular electronics.

Transport measurements through individual molecules or molecular assemblies, using either a STM or a break junction configuration, have demonstrated the characteristic zero bias conductance resonance indicative of the Kondo effect.  This resonance occurs due to an exchange process that results in the screening of the local magnetic moment, represented by an unpaired spin on the molecule, by the sea of delocalized conduction electrons.  Single molecule devices represent a platform for probing these interactions at more accessible temperatures compared to semiconductor QDs.  Furthermore, the ability to study transport in this regime using a variety of different systems (\emph{e.g.} QDs, carbon nanotubes, single molecules) aids researchers in distinguishing between universal behaviors and system related influences.  Transport measurements in the Kondo regime of SMTs can exhibit characteristics not seen in their QD counterparts and not fully accounted for by commonly used theoretical models.  This highlights the need for improved modeling techniques capable of incorporating complex correlations in a realistic molecular system.  Additionally, it illustrates the necessity for elucidating the unique qualities of the single molecule system, such as the precise way in which molecular orbital levels are renormalized when a molecule is coupled to metallic electrodes.  Many open question remain in the study of Kondo physics in molecular electronic devices, leaving much room for future investigations of this complex phenomenon.

DN acknowledges support from the David and Lucille Packard Foundation,
Robert A. Welch Foundation grant C-1636, NSF DMR-0347253 and
DMR-0855607.  GDS acknowledges the support of the W. M. Keck Program
in Quantum Materials at Rice University.

\end{document}